# Major depression as a complex dynamic system

Angélique O.J. Cramer[1*], Claudia D. van Borkulo[1], Erik J. Giltay[2], Han L.J. van der Maas[1], Kenneth S. Kendler[3], Marten Scheffer[4], Denny Borsboom[1]

[1] Psychological Methods, University of Amsterdam

[2] Department of Psychiatry, Leids Universitair Medisch Centrum

[3] Virginia Institute of Psychiatric and Behavioral Genetics, Virginia Commonwealth University

[4] Department of Aquatic Ecology, Wageningen University

\* Corresponding author

E-mail: aoj.cramer@gmail.com (AOJC)

# Abstract

In this paper, we characterize major depression (MD) as a complex dynamic system in which symptoms (e.g., insomnia and fatigue) are directly connected to one another in a network structure. We hypothesize that individuals can be characterized by their own network with unique architecture and resulting dynamics. With respect to architecture, we show that individuals vulnerable to developing MD are those with strong connections between symptoms: e.g., only one night of poor sleep suffices to make a particular person feel tired. Such vulnerable networks, when pushed by forces external to the system such as stress, are more likely to end up in a depressed state; whereas networks with weaker connections tend to remain in or return to a non-depressed state. We show this with a simulation in which we model the probability of a symptom becoming 'active' as a logistic function of the activity of its neighboring symptoms. Additionally, we show that this model potentially explains some well-known empirical phenomena such as spontaneous recovery as well as accommodates existing theories about the various subtypes of MD. To our knowledge, we offer the first intra-individual, symptom-based, process model with the potential to explain the pathogenesis and maintenance of major depression.

# Introduction

Major depression (MD) imposes a heavy burden on people suffering from it. Not only are the symptoms of MD themselves debilitating, their potential consequences (e.g., stigmatization and interpersonal rejection) can be equally detrimental to long-term physical and mental health [1]-[4]. Combined with the fact that MD approximately affects 17% of the general population at some point in their lives, denoting MD as one of the biggest mental health hazards of our time is hardly an overstatement [5]-[7]. It is therefore surprising, and somewhat disappointing, that we still have not come much closer to unraveling the pathogenesis of MD: what makes some people vulnerable to developing MD? The main aim of the present paper is to investigate this general question about MD from a network perspective on psychopathology, by means of developing a formal dynamic systems model of MD and conducting two simulation studies based on this model.

## What is MD as a complex dynamic system?

The *network perspective* on mental disorders comprises a relatively new branch of theoretical and statistical models [8]-[12]. Although the basic idea of networks is not new (e.g., see [13]-[18]), current network models extend this earlier theoretical work with a coherent framework hypothesized to deliver a blueprint for the development of a multitude of mental disorders [8]-[12]. Additionally, the network perspective currently comprises a number of methods to estimate and analyze such networks. The network perspective on psychopathology starts out by assuming that symptoms (e.g., MD symptoms such as trouble sleeping, fatigue, and concentration problems) cause other symptoms. For example, after an extended period of time during which a person is suffering from insomnia, it is not surprising that this person will start experiencing fatigue: insomnia → fatigue. Subsequently, if the fatigue is longer lasting, this person might start developing concentration problems: fatigue → concentration problems.

According to the network perspective, such direct relations between MD symptoms have, *theoretically speaking*, the capacity to trigger a diagnostically valid episode of MD: insomnia → fatigue → concentration problems → depressed mood → feelings of self-reproach, resulting in five symptoms on the basis of which a person is diagnosed with an episode of MD.

MD as such a network of directly related symptoms is more generally referred to as a *complex dynamic system* [19]: *complex* because symptom-symptom relations might result in outcomes, an episode of MD for instance, that are impossible to predict from any individual symptom alone; *dynamic* because this network of symptom-symptom relations is hypothesized to evolve in an individual over time; and a *system* because the pathogenesis of MD is hypothesized to involve direct relations between symptoms that are part of the same system. MD specifically is hypothesized to be a *bistable* system with two *attractor* states: a 'non-depressed' and a 'depressed' state.

## Aim of this paper

Evidence in favor of the network perspective is accumulating [20]. The current state of affairs can be summarized as follows: we know with a reasonable degree of certainty that symptom-symptom relations are present in groups of individuals, but we do not know *what* makes such symptom-symptom relations of an individual patient with MD different from the very same symptom-symptom relations of someone without MD. In other words, what makes the networks of some individuals more *vulnerable* to develop an episode of MD compared to networks of individuals who will not/never develop such an episode? Answering this question takes the dynamic systems perspective on MD the next critical steps further and is therefore the main goal of this paper. So what is vulnerability in the MD dynamic system?

## Vulnerability in the MD dynamic system

The generic diathesis-stress model [21]-[25] attempts to answer questions such as why some people develop MD after experiencing stressful life events while others do not. Derivatives of this general model have in common the hypothesis that developing a disorder such as MD is the result of an interaction between a certain *diathesis* (i.e., vulnerability) and a range of possible stressors. More specifically, the experience of a certain stressful life event can activate the diathesis (e.g., [26]).

But what is the diathesis, what is it that makes certain people vulnerable? Quite a few theories attempt to answer this question (e.g., certain risk alleles, high level of neuroticism; [27]-[30]) but, in this paper, we propose an alternative. This alternative is based on the notion that individuals likely differ, among other things, in terms of how strong certain symptoms are connected in their networks. For example, Carol has to suffer from at least four consecutive sleepless nights before she starts experiencing fatigue (i.e., a relatively weak connection between insomnia and fatigue) while Tim feels fatigued after only one sleepless night (i.e., a relatively strong connection between insomnia and fatigue). Now, we hypothesize that one of the ways in which a network is vulnerable to developing an episode of MD is the presence of strong connections between symptoms.

Vulnerability in a network is perhaps best illustrated by considering the symptoms of an MD network to be domino tiles and regarding the connections between them as the distances between the domino tiles [9]. Fig 1 shows this analogy. Strong connections (i.e., a vulnerable network) are analogous to domino tiles that stand in close proximity to one another (right panel of Fig 1): if one symptom becomes active in such a vulnerable network then it is highly likely that this activated symptom will result in the development of other symptoms. That is, in the analogy, the toppling of one domino tile will topple the other dominoes because the distances between them are short. On the other hand, weak connections (i.e., an *invulnerable* network) are analogous to domino tiles that are widely spaced (left panel of Fig

1): the development of one symptom is not likely to set off a cascade of symptom development because the symptom-symptom relations are not strong. That is, in the analogy, the toppling of one domino tile will not likely result in the toppling of others because of the relatively large distances between them.

**Fig 1. An analogy between vulnerability in a network and spacing of domino tiles.**

We developed the vulnerability hypothesis based on three general observations: 1) network models from other areas of science show that strong connections between elements of a dynamic system predict the tipping of that same system from one attractor state into another [31], [32]; 2) quite a few successful therapeutic interventions specifically aim to weaken or eliminate symptom-symptom relations (e.g., exposure therapy that aims at breaking the connection between seeing a spider and responding to it with fear by repeatedly exposing a patient to (real) spiders; [33], [34]); 3) increasing evidence that various patient groups have stronger network connections between psychopathological variables compared to healthy controls or patient groups in remission [35]-[37]. However, due to the cross-sectional nature of these data, it remains thus far an open question if these results readily generalize to intra-individual networks.

In the next section we introduce our formal network model of MD. This formal model will be the starting point of a simulation study that will be conducted in two parts. In the first simulation (Simulation I), we exclusively investigate the influence of increasing connectivity (i.e., diathesis or vulnerability) on the behavior of an MD system. The main question here is if a system with stronger connections will end up in a depressed state more easily than a system with relatively weak connections. In the second simulation (Simulation II) we examine the influence of stress. Here, the main question is what happens if we put vulnerable networks under stress.

# Simulation I: Investigating the vulnerability hypothesis

In this section, we build a formal dynamic systems model of MD in two steps (please see Fig 2 for a visualization of these steps). In the first step, we estimate threshold and weight parameters for an empirical inter-individual network of MD symptoms based on empirical data (see Fig 2A). In the second step, with these empirical parameters, we build a dynamic intra-individual model of MD which develops over time (see Fig 2B). The main characteristic of the model is that the activation of a symptom influences the probability of activation of other symptoms in its vicinity. We simulate data with this model in order to test the hypothesis that strongly connected MD networks are more vulnerable to developing a depressed state than weakly connected MD networks.

**Fig 2. A visualization of the setup of Simulation I.**

Panel A features a simplified network for variables $X1 - X9$ of the VATSPUD data. From these data we estimated weight parameters (i.e., the lines between the symptoms: the thicker the line the stronger the connection) and thresholds (i.e., the filling of each node: the more filling the higher the threshold). These empirical parameters were entered into the simulation model (black and red dashed arrows from panel A to panel B). To create three MD systems, we multiplied the empirical weight parameters with a connectivity parameter $c$ to create a system with weak, medium and strong connectivity. Panel B shows a gist of the actual simulation: for the three MD systems, we simulated 1000 time points (with the equations given in the main text) and at each time point, we tracked symptom activation. Our goal was to investigate our hypothesis (most right part of panel B) that the system with strong connectivity would be the most vulnerable system, i.e., with the most symptoms active over time.

# Methods

## VATSPUD data set

The Virginia Adult Twin Study of Psychiatric and Substance Use Disorders (VATSPUD) is a population-based longitudinal study of 8973 Caucasian twins from the Mid-Atlantic Twin Registry ([38], [39]). The first VATSPUD interview – the data of which were used for this paper – assessed the presence/absence of the 14 disaggregated symptoms of MD (representing the nine aggregated symptoms of criterion A for MD in DSM-III-R), lasting at least 5 days during the previous year (i.e., the data is binary). Whenever a symptom was present, interviewers probed to ensure that its occurrence was not due to medication or physical illness. Co-occurrence of these symptoms during the previous year was explicitly confirmed with respondents. The sample contained both depressed and non-depressed respondents (prevalence of previous year MD was 11.31%).

## Deriving empirical parameters

We estimated network parameters for the 14 symptoms of the VATSPUD dataset with a recently developed method, based on the Ising model, which reliably retrieves network parameters for binary data with good to excellent specificity and sensitivity. The model is easy to use as it is implemented in the freely available *R*-package *IsingFit* [40]. With this method, one estimates two sets of parameters [41]: 1) *thresholds*: each symptom has a threshold $\tau_i$ which is the extent to which a symptom *i* has a preference to be 'on' or 'off'. A threshold of 0 corresponds to a symptom having no preference while a threshold of higher (lower) than 0 corresponds to a symptom with a preference for being 'on' ('off'). In Fig 2 a threshold is visualized as a red filling of the nodes: the more the node is filled, the higher the threshold, which corresponds to a preference of that node to be 'on'. Less filling of a node corresponds with a lower threshold, which corresponds to a preference of that node to be 'off'; 2) *weights*: a weight $w_{ij}$ corresponds to a pairwise connection between two symptoms $i$ and $j$; if $w_{ij} = 0$ there is no connection between symptoms $i$ and $j$. The higher (lower) $w_{ij}$ becomes, the more symptoms $i$ and $j$ prefer to be in the same (different) state ('on' or 'off'). In Fig 2 a weight is visualized as a line (i.e., edge) between two nodes: the thicker the edge, the stronger the preference of these nodes to be in the same state ('on' or 'off'). Note that threshold and weight parameters are independent from one another. Both thresholds and weight parameters were estimated within a L1-regularized logistic regression model with an extended Bayesian Information Criterion (EBIC) as model selection criterion.

**The formal dynamic systems model of MD**

We begin developing the formal model of MD by assuming the following: 1) symptoms ($X_i$) can be 'on' (1; active) or 'off' (0; inactive); 2) symptom activation takes place over time ($t$) such that, for example, insomnia at time $t$ may cause activation of fatigue at time $t + 1$; and 3) a symptom $i$ receives input from symptoms with which it is connected in the VATSPUD data (i.e., these are non-zero weight parameters). These weight parameters are collected in a matrix **W** for the $J = 14$ symptoms: entry $W_{ij}$ thus represents the logistic regression weight between symptoms $i$ and $j$ as estimated from the VATSPUD data (as one can see in Fig 2 the weight parameters from the data are used in the subsequent simulations with our model).

Model formulation now proceeds along the following steps:

- We assume that the total amount of activation a symptom $i$ receives at time $t$ is the weighted (by **W**) summation of all the neighboring symptoms **X** (i.e., the vector that contains the "0" and "1" values of being inactive and active respectively) at time $t - 1$. We call this the *total activation function* (boldfaced parameters are estimated from the VATSPUD data):

$$A_i^t = \sum_{j=1}^{J} \boldsymbol{W}_{ij} X_j^{t-1}$$

- We formulate a logistic function for computing the probability of symptom *i* becoming active at time *t*: the probability of symptom *i* becoming active at time *t* depends on the difference between the total activation of its neighboring symptoms and the threshold of symptom *i* (in the formula below: ($\boldsymbol{b_i}$ - $A_i^t$)). This threshold is estimated from the VATSPUD data (see also Fig 2). Note that the parameter $\boldsymbol{b_i}$ denotes the absolute value of these estimated thresholds. The more the total activation exceeds the threshold of symptom *i* at time *t*, the higher the probability that symptom *i* becomes active (in the formula below: $P(X_i^t = 1)$) at time *t*. We call this the *probability function* (boldfaced parameters are estimated from the VATSPUD data):

$$P(X_i^t = 1) = \frac{1}{1 + e^{(\boldsymbol{b_i} - A_i^t)}}$$

To summarize: our model is an intra-individual model that develops over time. The probability of a symptom becoming active at a particular point in time depends on both its threshold and the amount of activation it receives from its neighboring symptoms at that same point in time. The more activation a symptom *i* receives from its neighboring symptoms and the lower its threshold, the higher the probability of symptom *i* becoming active.

**The simulation study**

To investigate our vulnerability hypothesis, we inserted a connectivity parameter *c* with which matrix **W** is multiplied. This results in the following modified *total activation function*:

$$A_i^t = \sum_{j=1}^{J} c\boldsymbol{W}_{ij} X_j^{t-1}$$

This connectivity parameter $c$ took on three values to create three networks (see also Fig 2B for a visualization of the simulation): 1) weak ($c = 0.80$); 2) medium ($c = 1.10$); and 3) strong connectivity ($c = 2.00$). For all three networks, we simulated 10000 time points starting with all symptoms being 'off' (i.e., **X** vector with only zeroes). At each time point, we computed total activation and the resulting probability of a symptom becoming active. Next, symptom values (either "0" or "1", denoting inactive and active, respectively) were sampled using these probabilities. Subsequently, at each time point, we tracked the state of the entire system, $D$, by computing the total number of activated symptoms (i.e., $D = \Sigma(X)$): the more symptoms are active at time $t$, the higher $D$ and thus the more 'depressed' the system is. The minimum value of $D$ at any point in time is 0 (no symptoms active) while the maximum value is 14 (all symptoms are active). We predicted that the network with the strongest connectivity (i.e., the highest weight parameters) would, over time, show the highest levels of $D$ compared to the networks with medium and weak connectivity (in Fig 2: the blue bar ranging from light blue for the network with weak connectivity (few symptoms; invulnerable) to dark blue for the network with strong connectivity (many symptoms; vulnerable).

## Results and Discussion

Fig 3 presents the network that resulted from the parameter estimation with *IsingFit*. The edges between the symptom nodes represent the estimated logistic regression weights (note: thresholds are not visualized in this network but are given in the right panel next to the network). The positioning of the nodes is such that nodes with strong connections to other nodes are placed towards the middle of the network. Nodes with relatively weak connections to other nodes are placed towards the periphery of the network.

**Fig 3. The inter-individual MD symptom network based on the VATSPUD data.** Each node in the left panel of the figure represents one of the 14 disaggregated symptoms of MD

according to DSM-III-R. A line (i.e., edge) between any two nodes represents a logistic regression weight: the line is green when that weight is positive, and red when negative. An edge becomes thicker as the regression weight becomes larger. As an example, the grey circles are the *neighbor* of the symptom that is encircled in purple (i.e., they have a connection with the purple symptom). The right part of the figure shows the estimated thresholds for each symptom. *dep*: depressed mood; *int*: loss of interest; *los*: weight loss; *gai*: weight gain; *dap*: decreased appetite; *iap*: increased appetite; *iso*: insomnia; *hso*: hypersomnia; *ret*: psychomotor retardation; *agi*: psychomotor agitation; *fat*: fatigue; *wor*: feelings of worthlessness; *con*: concentration problems; *dea*: thoughts of death.

The results of the simulation study for the first 1500 time points are presented in Fig 4. As we predicted, the stronger the connections in the MD system, the more vulnerable the system is for developing depressive symptoms (as tracked with the symptom sum score, or state, $D$ at each time $t$): in the weakly connected system (most left graph at the top of Fig 4) there certainly is some development of symptoms (i.e., peaks in the graph) but the system never quite reaches a state $D$ where many symptoms are developed. As one can see in this graph, the symptom sum score $D$ is nowhere higher than 7. In the case of medium connectivity (middle graph at the top of Fig 4) the system is capable of developing more symptoms (i.e., higher values of $D$, peaks in the graph) compared to the weak connectivity network. On the other hand, that same system returns (quite rapidly) to non-depressed states (i.e., lower values of $D$, dips in the graph). The strong connectivity system (most right graph at the top of Fig 4) is clearly the most vulnerable: the system settles into a depressed state rapidly (i.e., high and sometimes maximum values of $D$) and never exits this state.

**Fig 4. The results of Simulation I.**

The top of the figure displays three graphs: in each graph, the state of the system *D* (i.e., the total number of active symptoms; *y*-axis) is plotted over time (the *x*-axis). From left to right, the results are displayed for a weakly, medium and strongly connected network respectively. For the network with weak connections, we zoom in on one particular part of the graph in which spontaneous recovery is evident: there is a peak of symptom development and these symptoms spontaneously become deactivated (i.e., without any change to the parameters of the system) within a relatively short period of time.

What stands out in the graph of the weakly connected MD system is the presence of *spontaneous recovery*. We zoomed in at one particular part of the time-series (see 'zoomed in' at the bottom of Fig 4) in which one can clearly see a point where 7 symptoms are active (right in the middle of the graph). Without any change to the parameters the system recovers spontaneously (and rapidly) to a state in which no symptoms are active (i.e., a non-depressed state, $D = 0$). To our knowledge, we are the first to show spontaneous recovery in a formal model of MD and as such, the results offer a testable hypothesis: spontaneous recovery is most likely to occur in people whose MD symptoms are not strongly connected.

One hypothesized subtype of MD is *endogenous* with bouts of depression that appear to come out of the blue, without any apparent external trigger such as a stressful life event (e.g., [42]). One could argue that this is exactly what happens in our simulation of a strongly connected MD system. There are no external influences on the system and the parameters of the system (e.g., thresholds, weights) remain the same throughout the 10000 simulated time points. Yet in the strongly connected network, the development of only one symptom is apparently enough to trigger a cascade of symptom development with a depressed state of the system as a result (most right graph at the top of Fig 4). As such, endogenous depression might be characterized as strong connections in someone's MD system but due to the exploratory nature of this finding, confirmatory studies are needed before any definitive conclusions can be drawn.

## Simulation II: Investigating the influence of external stress on MD systems

In Simulation I we studied vulnerability in isolation, that is, without any external influences on the MD system. While insightful such a model does not do justice to the well-established fact that external pressures such as stressful life events (e.g., the death of a spouse) have the potential to – in interaction with vulnerability – cause episodes of MD (i.e., diathesis-stress models as we outlined earlier; [43]-[47]). In fact, this non-melancholic subtype for which an episode can be partially explained by environmental circumstances, is quite prevalent. Therefore, the aim of this section is to investigate the interaction between our conceptualization of vulnerability as established in Simulation I (i.e., the diathesis of a strongly connected symptom network) and stress: what happens if we put stress on a system with increasing connectivity (i.e., higher weight parameters)?

More specifically, we will investigate what happens within the context of the cusp catastrophe model. One of the problems with networks is that they easily become very complex. Even our relatively simple model with 14 symptoms already entails more than 100 parameters (14 thresholds and 91 weight parameters). When adding other parameters (e.g., a stress parameter) the model quickly becomes more intractable and as such less informative about the general behavior of the system. It is therefore customary in other fields (e.g., the dynamics of the coordination of certain movements; [48]) to use the *cusp catastrophe model* as a way of simplifying the model just enough in order to understand its general dynamics [49]-[54]. The cusp catastrophe model is a mathematical model that can explain why small changes in some parameter (in our model: a small increment in external stress) can result in catastrophic changes in the state of a system (in our model: a shift from a non-depressed to a depressed state, or vice versa). The cusp catastrophe model (see Fig 5 for a visualization of this model) uses two orthogonal control variables, the *normal* variable (i.e., the *x*-axis) and the *splitting* variable (i.e., the *y*-axis) that, together, predict behavior of a given system (i.e., the *z*-axis). We hypothesize that stress acts as a normal variable while connectivity acts as the splitting variable.

**Fig 5. A visualization of a cusp catastrophe model.**

This figure features two panels: (A) The 3D cusp catastrophe model with stress on the *x*-axis, connectivity on the *y*-axis and the state of the system (i.e., *D*: the total number of active symptoms) on the *z*-axis; and (B) A 2D visualization of the cusp as depicted in (A). In the case of weak connectivity (top graph in (B)), the system shows smooth continuous behavior in response to increasing stress (green line, invulnerable networks). In the case of strong connectivity (bottom graph in (B)), the system shows discontinuous behavior with sudden jumps from non-depressed to more depressed states and vice versa (red line, vulnerable networks). Additionally, the system with strong connectivity shows two tipping points with in between a so-called forbidden zone (i.e., the dashed part of the red line): in that zone, the state of the system is unstable to such an extent that even a minor perturbation will force the system out of that state into a stable state (i.e., the solid parts of the red line).

What are the main characteristics of this model?
- With increasing values of the splitting variable (i.e., connectivity) the behavior of the system becomes increasingly *discontinuous*. In Fig 5B (a 2D representation of Fig 5A): as stress increases but connectivity is weak (top graph of Fig 5B; invulnerable networks), the solid green line shows that the state of the system becomes more 'depressed' in a smooth and continuous fashion. To the contrary, as stress increases but connectivity is strong (bottom graph of Fig 5B; vulnerable networks), the red line shows that the state of the system becomes more 'depressed' in a discontinuous fashion.

- For vulnerable networks one should expect to see two *tipping points* between which a so-called 'forbidden' zone is present (in bottom graph of Fig 5B: the part of the red line that is dashed): within this zone, the state of the system is unstable to such an extent that even a very modest disturbance (e.g. a mild stressor) may already kick the system out of equilibrium into more depressed states. Such tipping points are preceded by early warning signals, most notably *critical slowing down* [55]-[61]: right before a tipping point, the system is becoming increasingly slower in recovering (e.g., person remains sad and sleeps badly for a prolonged time) from small perturbations (e.g., a minor dispute).

- *Hysteresis* for vulnerable networks: once the MD system has gone through a catastrophic shift to an alternative state (e.g., person becomes depressed), it tends to remain in that new state until the external input (i.e., stress) is changed back to a much lower level than was needed to trigger that depressed state (e.g., solving marital problems that triggered an episode of MD will not be sufficient to get that person into a non-depressed state).

We use this model in this section in three ways: 1) we check to what extent the results of the simulations match with the characteristics of a cusp catastrophe model; 2) we directly test the hypothesis that stress acts as a normal variable while connectivity acts as the splitting variable; and 3) we investigate potential early warnings of upcoming transitions from one state into another, a prediction that follows from a cusp catastrophe model.

# Methods

## The formal dynamic systems model of MD

For the sake of simplicity, we assumed that stress influenced all symptoms in an equal manner (see left part of Fig 6, a visualization of the setup of Simulation II). To this end, we extended our formal model of MD – see *Methods* of Simulation I – with a stress parameter $S_i^t$, a number that was added to the total activation of the neighbors of symptom *i* at time *t*: the higher $S_i^t$ – that is, the more stress – the higher the total activation function, and thus the higher the probability that symptom *i* will become active at time *t*. This results in the following modified *total activation function*:

$$A_i^t = \sum_{j=1}^{J} c W_{ij} X_j^{t-1} + S_i^t$$

**Fig 6. A visualization of the setup of Simulation II.**

First, we put stress on all the symptoms of the systems with weak, medium and strong connectivity by adding a stress value to the total activation function of each symptom (left part of the figure). Then, we simulate 10000 time points during which we 1) increase and decrease stress and 2) track symptom activation at each time point (right part of the figure).

As a reminder, in this function *t* denotes time, *c* is the connectivity parameter that takes on three values: 1) weak connectivity (*c* = 0.80); 2) medium connectivity (*c* = 1.10); and 3) strong connectivity (*c* = 2.00). Matrix **W** encodes the weight parameters that were estimated from the VATSPUD data. Vector **X** contains the status of symptoms ("0", inactive, or "1", active) at the previous time point *t* – 1. The *probability function* remained equal to the one used in Simulation I.

## The simulation study

Analogous to Simulation II, we simulated 10000 time points for each of the three values of the connectivity parameter *c*. For these three types of systems, we observed the impact of variation in the stress parameter (see right part of Fig 6): over the course of the 10000 time points $S_i^t$ was repeatedly gradually increased from -15 to 15 and then decreased from 15 to -15 with small steps of 0.01 (the numerical values of the stress parameter and the steps were chosen randomly). The impact of the stress parameter on the behavior of the system was quantified by computing the *average* state *D* of the system, that is, the average number of symptoms active at a certain time point *t*. Specifically, since all the stress parameter values were used multiple times during the simulation – because of the increasing and decreasing of the stress parameter during the 10000 time points – we averaged states within 0.20 range of these stress parameter values. So for example, suppose that stress values between 9.80 – 10.20 come up 15 times during the 10000 simulated time points. Then, we computed the average state *D* by taking all states *D* within the 9.80 – 10.20 range of stress parameter values and dividing this sum score by 15.

## Fitting the cusp catastrophe model

We tested our hypothesis that stress acts as a normal variable while connectivity acts as the splitting variable with the *cusp* package in R [62]. With this package, one is able to compare different cusp models in which *S* (stress) and *c* (connectivity) load on none, one or on both control variables, very much in the same way in which test items load on factors in a factor model. For this test, we used the same simulation model as outlined above but we used a simple weights matrix **W** in which all weights were set to be equal.

## Critical slowing down

We quantified critical slowing down with *autocorrelations*: the correlations between values of the same variable at multiple time points. Such autocorrelations go up when the system slows down: slowing down means that at each time point, the system much resembles the system as it was at the previous time point, meaning that the autocorrelation is relatively high. We inspected the autocorrelations between the states *D* of the simulated vulnerable MD system at consecutive time points.

# Results and discussion of Simulation II

## Comparing simulation results to characteristics of cusp catastrophe model

Fig 7 shows the main results of the simulation: the *x*-axis represents stress while the *y*-axis represents the state of the system. The grey line (and points) represents the average number of active symptoms (for stress parameter values within 0.20 ranges) when stress was increasing; and the black line (and points) represents the average number of active symptoms when stress was decreasing. The figure shows that differences in network connectivity resulted in markedly different responses to external activation by stress. MD systems with weak connectivity proved invulnerable (left panel of Fig 7): stress increments led to a higher number of developed symptoms in a smooth continuous fashion, and stress reduction resulted in a smooth continuous decline of symptom activation. This is what one would expect to happen at the back of the cusp catastrophe model (see top graph Fig 5B). The dynamics were different for the systems with medium and strong connectivity (middle and right panel of Fig 7): as we expected from a cusp catastrophe model the behavior of the system became increasingly discontinuous as two tipping points appeared. That is, a small increase in stress could lead to a disproportional reaction, resulting in a more depressed state with more symptoms active. As such, we note here that, apparently, "…the hypotheses of kinds and continua are not mutually exclusive…" [63]: that is, our results show that, depending on connectivity, MD can be either viewed as a kind (in the case of a network with strong connectivity) or a continuum (in the case of a network with weak connectivity).

**Fig 7. The state of the MD system in response to stress for varying connectivity.**
The *x*-axis represents stress while the *y*-axis depicts the average state of the MD system, *D*: that is, the total number of active symptoms averaged over every 0.20 range of the stress parameter value. The grey line (and points) depicts the situation where stress is increasing (UP; from -15 to 15, with steps of 0.01) whereas the black line (and points) depicts the situation where stress is decreasing (DOWN; from 15 to -15, with steps of 0.01). The three graphs represent, from left to right, the simulation results for networks with low, medium, and high connectivity, respectively.

Additionally, and consistent with a cusp catastrophe model, both the medium and strong connectivity networks clearly showed that during the transition from non-depressed to more depressed states, or vice versa, a sizable 'forbidden zone' (from around 2 to 9 symptoms) was crossed that does not seem to function as a stable state (i.e., no data points in that area, see black boxes in Fig 7). Such a forbidden zone increases as a function of increasing connectivity. Therefore, the weak connectivity network (most left graph of Fig 7) shows a very small forbidden zone.

As was expected to happen at the front of the cusp catastrophe model (see Fig 5A), the results for the strong connectivity MD system showed clear hysteresis: the amount of stress reduction needed to get the system into a non-depressed state (i.e., only a few symptoms active or none at all) exceeds the amount of stress that tipped the system into depressed states in the first place. We checked for the robustness of the hysteresis effect by systematically repeating the simulation for different values of four parameters: 1) weights $W_{ij}$; 2) connectivity parameter *c*; number of nodes *J*; and 4) the $b_i$ parameter. Based on the results we conclude that the hysteresis effect is robust in that increasing connectivity of a network results in more hysteresis.

We are not aware of other (simulation) studies that showed hysteresis in MD symptom networks that are vulnerable to developing episodes of MD. The results do seem to resonate with clinical observations concerning the non-linear course of affective shifts between non-depressed and depressed states that is frequently encountered in the empirical literature [64].

**Fitting the cusp catastrophe model**

The best fitting model was the one in which only $c$ loaded on the splitting variable – as we hypothesized – but both $S$ and $c$ loaded onto the normal variable (for **W** with relatively small positive weights). As such, the normal and splitting axes are not strictly orthogonal and we take this to mean that our original mapping of the network dynamics require a nuance. An increase in connectivity has two effects in the cusp: it increased both the probability of more depressed states – because connectivity is part of the normal variable – and the hysteresis effect – because connectivity is also the splitting variable.

**Critical slowing down**

Fig 8 presents the results: as expected, when stress was increasing, the autocorrelations between the states of the MD system increase (dashed line increasing, starting at roughly the 0 stress point) before system abruptly switches from a non-depressed to a depressed state (thicker dashed line jumping from 0 to 14 symptoms, at roughly the 2 stress point). Additionally, when stress was decreasing, the autocorrelations increased as well (solid line increasing, starting roughly at the -2 stress point) before the system abruptly switches from a depressed to a non-depressed state (thicker solid line jumping from 0 to 14 symptoms, at roughly the -4 stress point).

Our results show that autocorrelations between the states of a system over time might provide a gateway into the prediction of tipping points. A recent empirical paper found similar increasing autocorrelations before a catastrophic shift in the time series of a single patient with MD [65]. Finding these tipping points for networks of actual, individual people could prove beneficial for two reasons. First, knowing that someone's MD system is close to tipping from a non-depressed to a depressed state would allow for precisely timed therapeutic interventions that might *prevent* such a catastrophic shift. Second, knowing that someone's MD system is close to tipping from a depressed to a healthy state would offer the opportunity of giving the system a large kick (e.g., electroconvulsive therapy) *at exactly the right time* so that the system is abruptly kicked out of a depressed state into a non-depressed state. Hence, knowing the tipping points of an individual's network might help in predicting when prevention and intervention have highest probability of success.

**Fig 8. Increasing autocorrelation as an early warning signal in the MD system with strong connectivity.**

The *x*-axis represents stress while the *y*-axis represents the average state: that is, the total number of active symptoms averaged over every 0.20 range of the stress parameter value. The dashed lines depict the situation where stress is increasing whereas the solid lines depict the situation where stress is decreasing. The "jump" lines show the total number of active symptoms (i.e., state), the "autocorrelation" lines track the autocorrelation between these states over time.

# Discussion

Throughout this paper we have advocated a view in which direct relations between symptoms have a crucial role in the pathogenesis of major depression (MD). We have developed a formal dynamic systems model of MD that was partly based on empirical data. We have conducted two simulation studies with the following resulting highlights: 1) strongly connected MD systems are most vulnerable to ending up in a depressed state; 2) putting vulnerable networks under stress results in discontinuous behavior with tipping points and hysteresis (consistent with a cusp catastrophe model); and 3) these vulnerable networks display early warning signals right before they tip into a (non-)depressed state. As such, we offer, to our knowledge, the first intra-individual, symptom-based, process model with the potential to explain the pathogenesis and maintenance of major depression while simultaneously accommodating for well-known empirical facts such as spontaneous recovery.

Adopting a dynamic systems approach to MD with symptom-symptom relations as its hallmark has empirical ramifications. For example, we argue that it might help in understanding mechanisms of change during treatment. For quite a few existing therapeutic strategies that appear to be at least moderately successful, mechanisms of change are not completely understood (e.g., cognitive behavioral therapy, CBT; [66], [67]). The apparent success of CBT might be understood as an attempt at reducing strong connectivity (e.g., by challenging a patient's irrational assumptions) between certain symptoms (e.g., between depressed mood and suicidal thoughts), or even breaking the connections altogether. As another example, a treatment strategy implied by a dynamic systems perspective is applying a perturbation to the system itself, which 'kicks' the system out of the depressed state ([60], [61]). For example, one could push the activation of a symptom to such an extreme (e.g., sleep depriving MD patients with insomnia; [68]) that it forces behavior that will eventually result in the deactivation of that symptom and/or, due to strong connectivity, other MD symptoms.

It is likely that the dynamic systems model we presented reaches beyond MD. For example, evidence is mounting in favor of a network perspective for disorders such as autism [69], post-traumatic stress disorder [70], schizophrenia [71] and substance abuse [72]. As such, our dynamic model of MD might serve as a starting point for investigating these and other disorders to which it may apply: if one has an inter-individual symptom-based dataset with an adequate number of respondents and empirically realistic prevalence rates, our code (http://aojcramer.com) can be used to run the simulations that we have reported in this paper.

A question that naturally arises when portraying MD, or another mental disorder, as a network of connected symptoms is where these connections come from. What do they really mean in terms of actual biological/psychological processes within a person? Take for example a direct relation between insomnia and fatigue: it stands to reason that such a direct relation, defined at the *symptom* level might be shorthand for events that actually take place in underlying *biological* regulatory systems. Alternatively, a connection in a network model might be shorthand for some (psychological) moderator, for example rumination that possibly serves as a moderator of the connection between feeling blue and feelings of worthlessness. The short and honest answer to the question what connections in a network really mean is that we do not know with any certainty at this point. A connection between any two symptoms can mean many things and future network-oriented research will need to tease apart the biological and/or psychological underpinnings of network connections [73]. While this may seem to be an important drawback of network modeling of psychopathology in general, we note that we generated some well-known empirical features of MD without any information about the origins of the connections between the MD symptoms whatsoever. That is: understanding a disorder might not necessarily entail knowing all there is to know about the real-world equivalents of the parameters of a model.

This paper has some limitations. First of all, for the sake of simplicity there was no autocatalysis in our model. That is, self-loops between a symptom and itself were set to 0. It might, however, be theoretically feasible to assume that at least for some of the symptoms of MD autocatalysis is in fact true. For example, insomnia might lead to even more insomnia because of worrying about the difficulties in falling asleep. Second, we held the thresholds for each symptom constant. In reality it might be reasonable to assume that individuals in fact differ in these thresholds. If thresholds are indeed idiosyncratic then the worst case scenario – in terms of vulnerability – would be the combination of strong connections between symptoms (dominos standing closely together) and low thresholds (it takes little to topple one domino). Finally, a useful extension of our model could be to incorporate the possibility that connectivity changes within a person [74], [75]: for example, it may be defensible to argue that a connection between two symptoms becomes stronger as these two symptoms are more frequently active within the same timeframe within a person.

By no means do we claim to have presented a model that, without further ado, explains all there is to know about MD. It is, however, high time to start rethinking our conceptualization of mental disorders in general – and MD in particular – and to at least entertain the proposition that "symptoms, not syndromes [i.e., latent variables] are the way forward" [76].

# Figures

*Figure 1.*

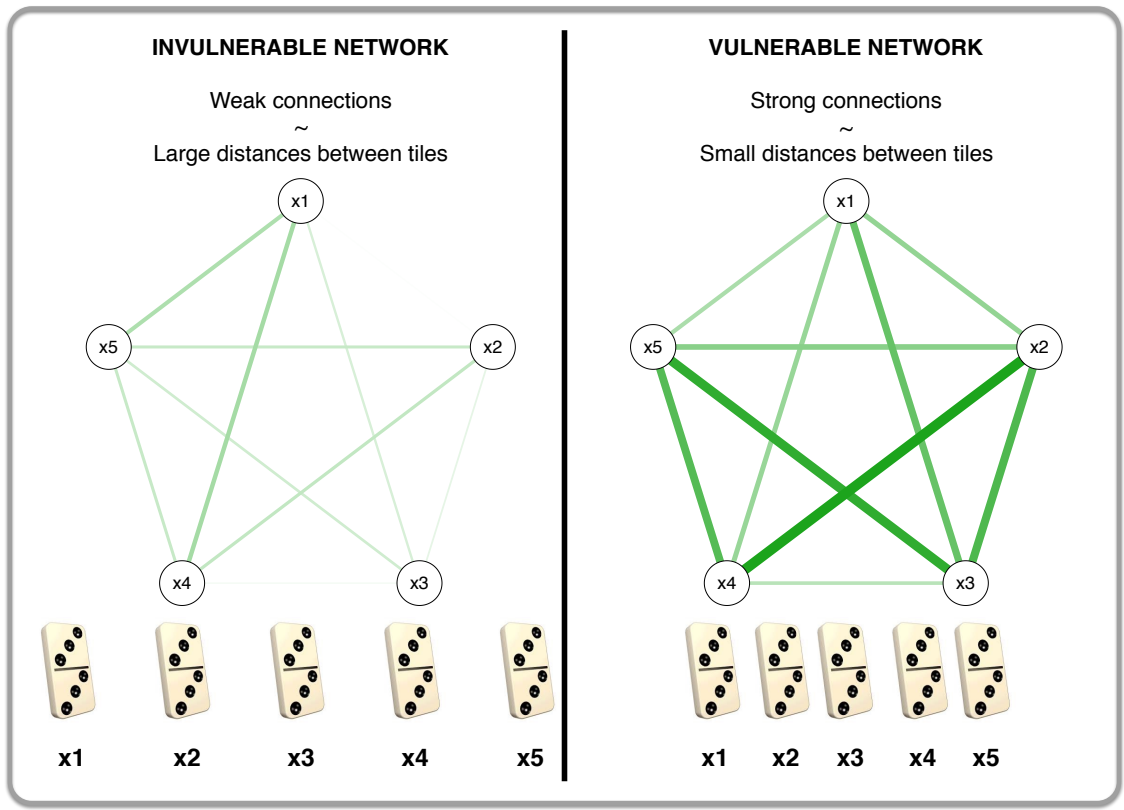

*Figure 2.*

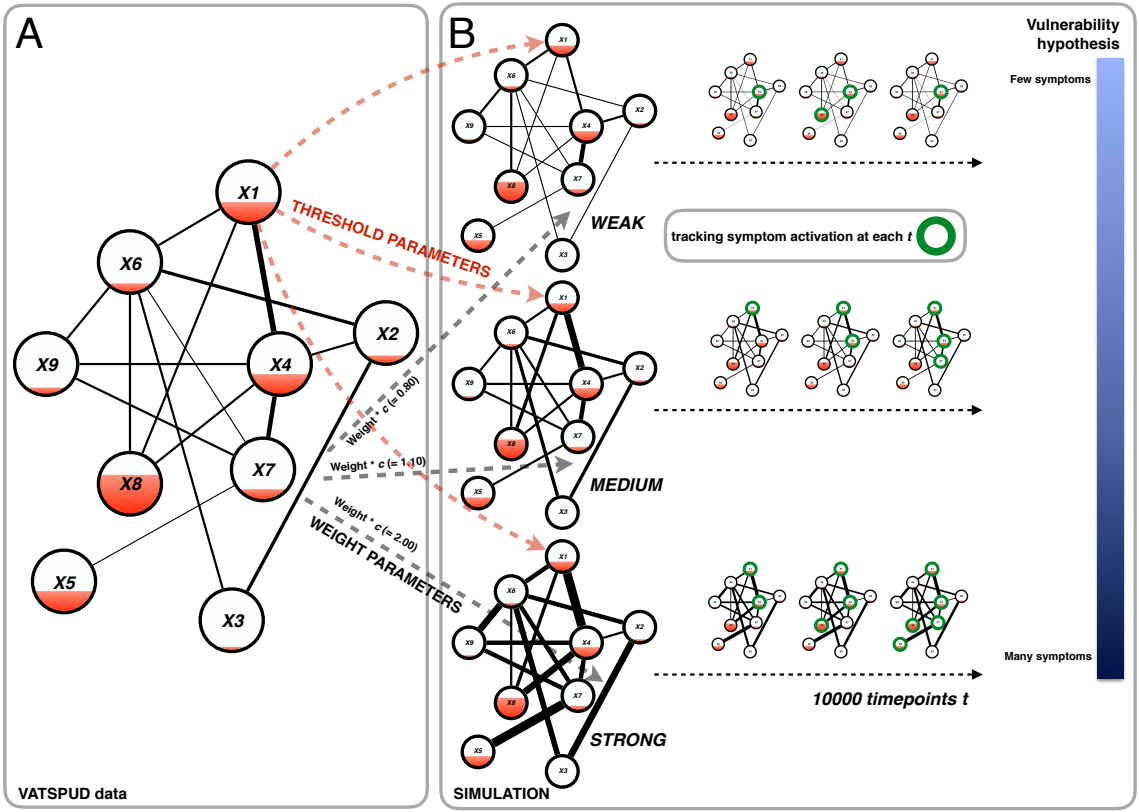

*Figure 3.*

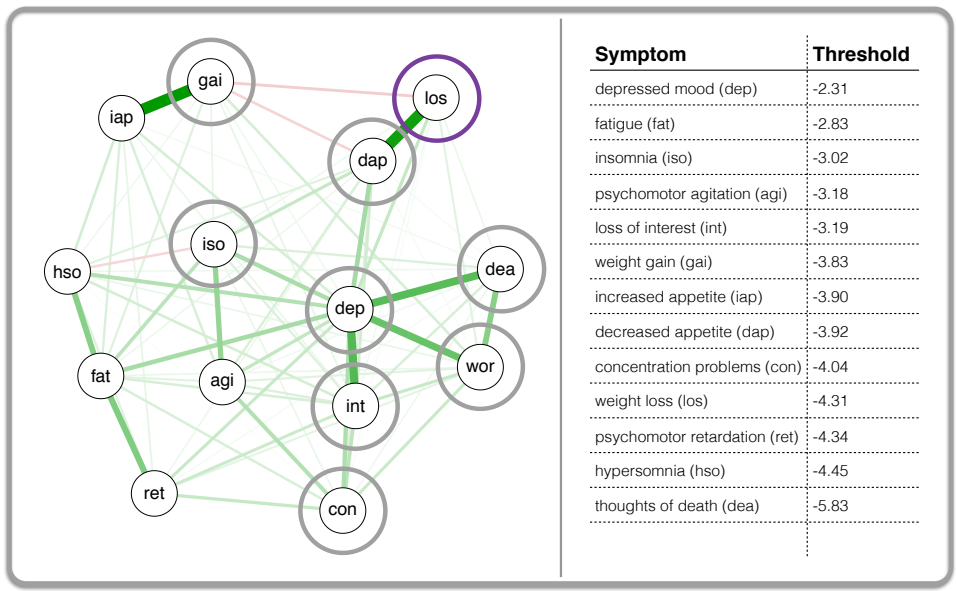

*Figure 4.*

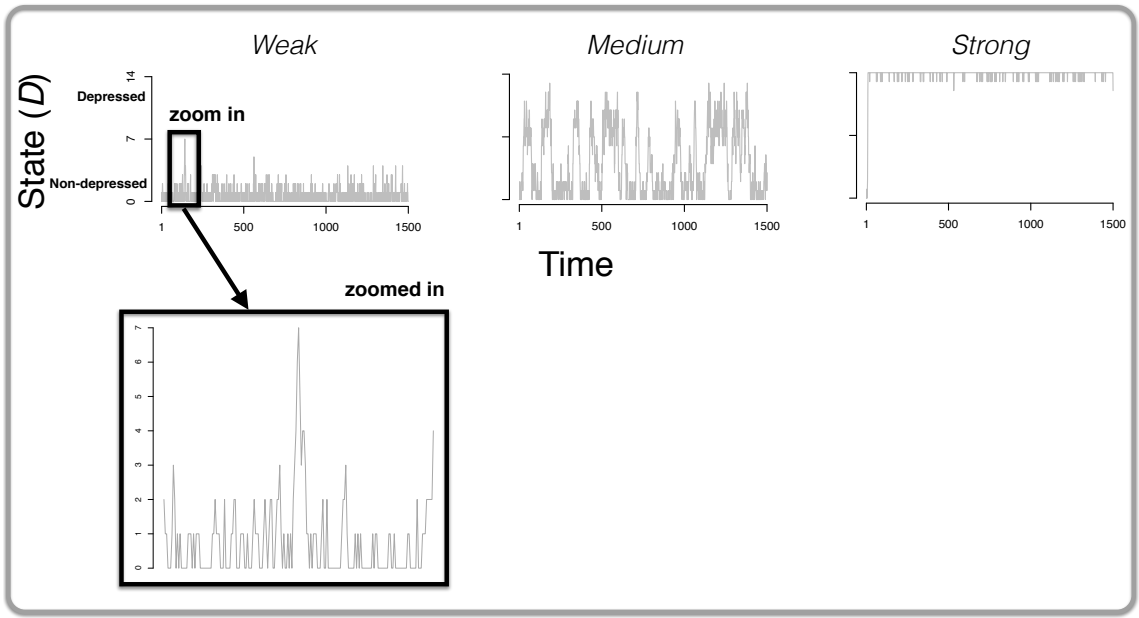

*Figure 5.*

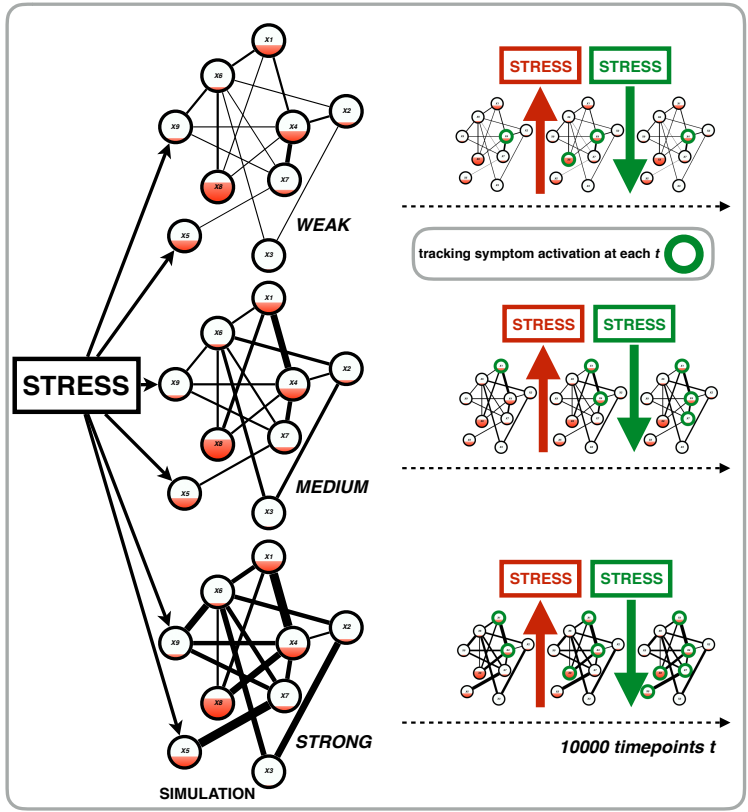

*Figure 6.*

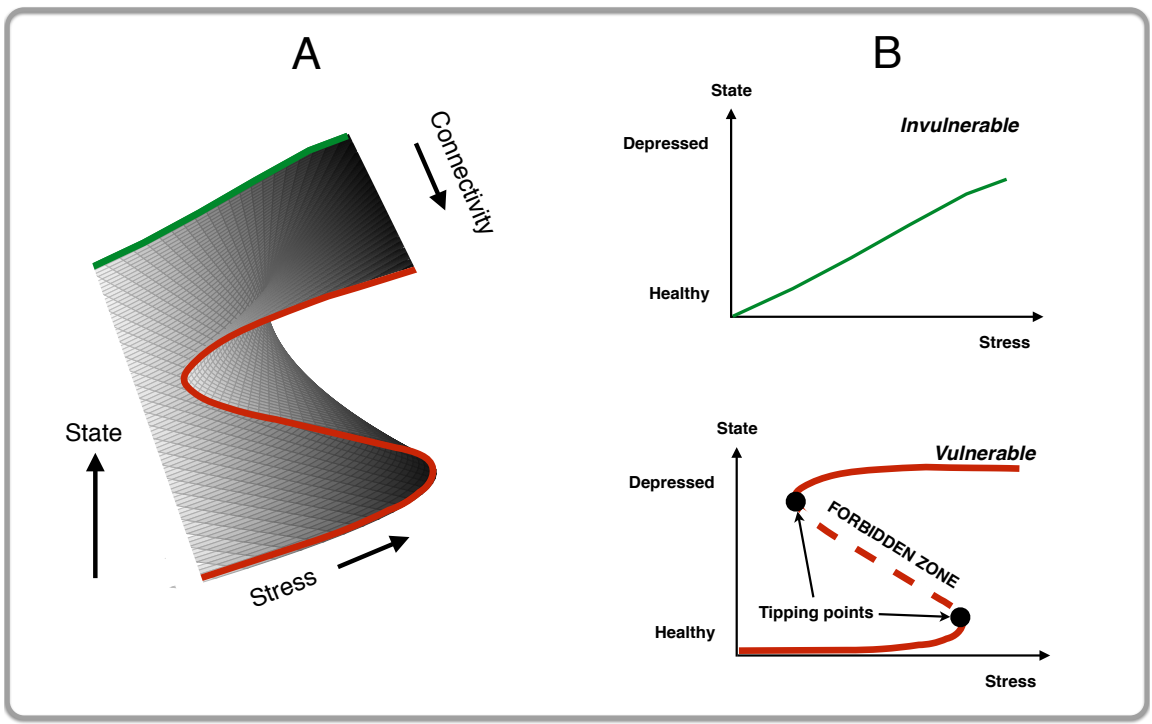

*Figure 7.*

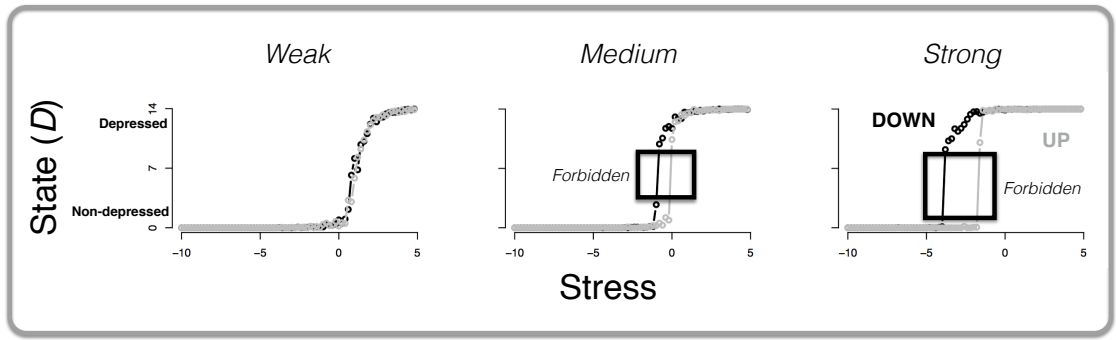

*Figure 8.*

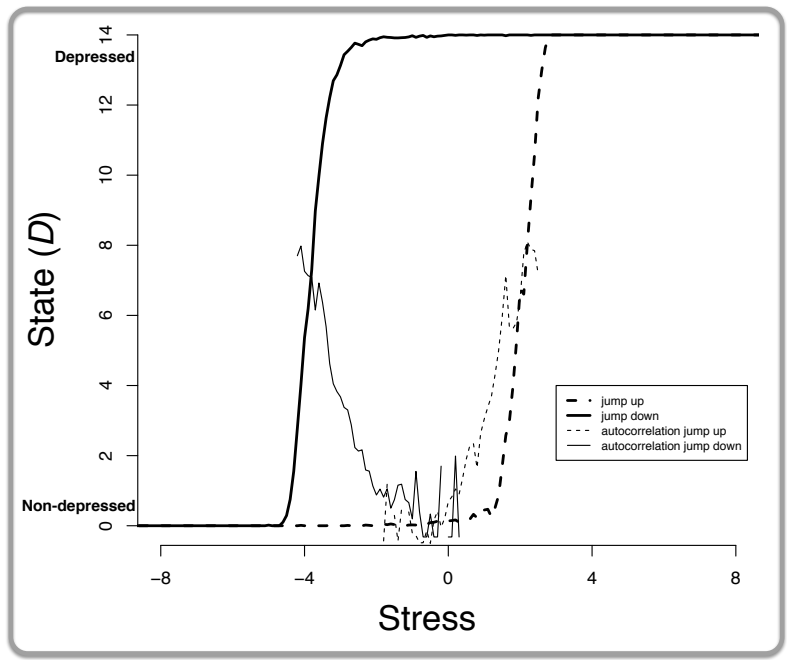